# Flat band excitons in a three-dimensional supertwisted spiral transition metal dichalcogenide

Yinan Dong[1,2]*, Yuzhou Zhao[3], Lennart Klebl[4, 5], Taketo Handa[6], Ding Xu[6], Chiara Trovatello[7], Chennan He[1], Dihao Sun[1], Thomas P. Darlington[7], Kevin W. C. Kwock[7], Jakhangirkhodja A. Tulyagankhodjaev[6], Yusong Bai[6], Yinming Shao[1,6], Matthew Fu[1], Raquel Queiroz[1], Milan Delor[6], P. James Schuck[7], Xiaoyang Zhu[6], Tim Wehling[5], Song Jin[3], Eugene Mele[9], Dmitri N. Basov*[1]

[1]Department of Physics, Columbia University, New York, NY, 10027, USA.

[2]Department of Applied Physics and Applied Mathematics, Columbia University, New York, NY, 10027, USA.

[3]Department of Chemistry, University of Wisconsin, Madison, WI, 53706, USA.

[4]Institut für Theoretische Physik und Astrophysik and Würzburg-Dresden, Cluster of Excellence ct.qmat, Universität Würzburg, 97074, Germany

[5]Institute of Theoretical Physics, University of Hamburg, Notkestraße 9-11, 22607 Hamburg, Germany.

[6]Department of Chemistry, Columbia University, New York, NY, 10027, USA.

[7]Department of Mechanical Engineering, Columbia University, New York, NY, 10027, USA.

[8]Department of Physics, Pennsylvania State University, University Park, PA, 16802, USA.

[9]Department of Physics, University of Pennsylvania, Philadelphia, PA, 19104, USA.

Corresponding authors* emails: yd2400@columbia.edu, db3056@columbia.edu

## Abstract

A frontier of van der Waals twistronics is the three-dimensional supertwisted material, where each successive atomic layer rotates the same angle towards the bulk limit. While two-dimensional moiré materials have been intensively studied, the unique consequences from 3D twistronics remain largely unexplored. Here we discover the flat band excitons in 3D supertwisted $WS_2$ by systematic photoluminescence experiments substantiated by electronic structure calculations. In addition to the photoluminescence feature resembling that of 2D A excitons, unique 3D direct and indirect exciton emissions are emergent and tied to the supertwist geometry. Both generalized Bloch band theory and local density of states calculations are constructed with screw rotational symmetry, revealing coexisting 2D and 3D flatband gaps that support these flatband excitons. The flatband excitons not only provide sensitive observables for the electronic properties of 3D supertwisted semiconductors but also unleash new pathways for quantum optoelectronics and topological polaritonics.

## Main

A twist between two van der Waals (vdW) material layers generates two-dimensional (2D) moiré effect; a continuous super-twist generates rotational super-moiré effect with three-dimensional (3D) properties. The former have unleashed a plethora of physical phenomena, such as superconductivity[1-3], ferroelectricity[4-6], and valleytronics[7]. Specifically, the semiconducting twisted transition metal dichalcogenides (TMDs) support bound electron-hole pairs, or excitons, which provide optical sensitivity to electronic properties[8,9] with the prominent example of moiré excitons[10- 23]. Building upon twisted bilayers, it is natural to increase the number of twisted layers for enhanced tunability and super-moiré effects. Towards the 3D limit, both in-plane Hamiltonian and interlayer electronic tunneling of the materials are governed by the underlying rotational symmetry, promising new topological physics[24]. However, challenges arise in both material fabrication and theoretical understanding as the system approaches the bulk limit. The commonly used tear-align-stack method becomes increasingly difficult to implement with more than a few layers. The lack of translational symmetry hinders the construction of conventional Bloch wavefunctions. Fortunately, advanced chemical vapor deposition technique has made the direct growth of 3D supertwisted TMDs[25,26] possible. Recent prediction of Weyl physics in 3D chiral semimetals[24] and demonstration of the opto-twistronic Hall effect[27] shed some light on the emerging physics. However, the full electronic and excitonic landscapes of supertwisted semiconductors remain unexplored, both theoretically and experimentally. In this study, we will bridge these gaps by systematically examining the excitonic properties of the 3D supertwisted WS$_2$ spirals originated from the flat bands, corroborated by electronic structure calculations.

## Structure of supertwisted spirals of TMDs

The electronic and excitonic properties of supertwisted TMDs are inherited from their unique geometric structure. Conventional TMDs exhibit crystalline structures including tetragonal (1T and distorted 1T') prototypes for monolayer TMDs and hexagonal (2H) and rhombohedral (3R) prototypes for multilayer TMDs (Fig. 1a). Figure 1b presents a supertwisted tetralayer WS$_2$ structure with an interlayer twist of 30° to illustrate the destruction of translational symmetry and appearance of a new screw rotational symmetry. As the number of layers increases, a supertwisted bulk is constructed where each of the atomic layers rotates the same angle (Fig. 1c). The supertwisted stacking features the absence of lattice translational symmetry and the presence of a global screw rotational symmetry. The actual crystal growth processes driven by central screw dislocation [25,26] result in a reduced lateral size for each successive layer and thus a macroscopic spiral pyramid structure, as shown in a top view sketch in Fig. 1d. Previous work has shown that deformation at the center of a supertwisted spiral can cause substantial strain[28]. To avoid this, we performed atomic force microscopy and scanning near-field optical microscopy (SNOM) to preselect structurally pristine 3D spiral samples and confirm the twist angles before conducting photoluminescence (PL) experiments. Figs. 1e and f present the topography of a typical non-twisted and a typical supertwisted spiral of WS$_2$, respectively, with corresponding optical scattering data imaged by a 745-nm laser, displayed in Figs. 1g and h. The subdiffractional periodic fringes in the SNOM channel are near-field waveguide polariton modes[29] indicating the material's high quality. The broken centrosymmetry of the material also leads to efficient nonlinear optical responses that are tied to the rotational structure. In the supplemental data, spatially resolved second harmonic generation (SHG) (Extended Data 1-2) and Raman experiments (Extended Data Fig. 3) show that the supertwist modulates the amplitude and polarization of signals deterministically. These nonlinear optical properties arise from the supertwisted structure and reflect the strong light-matter interaction.

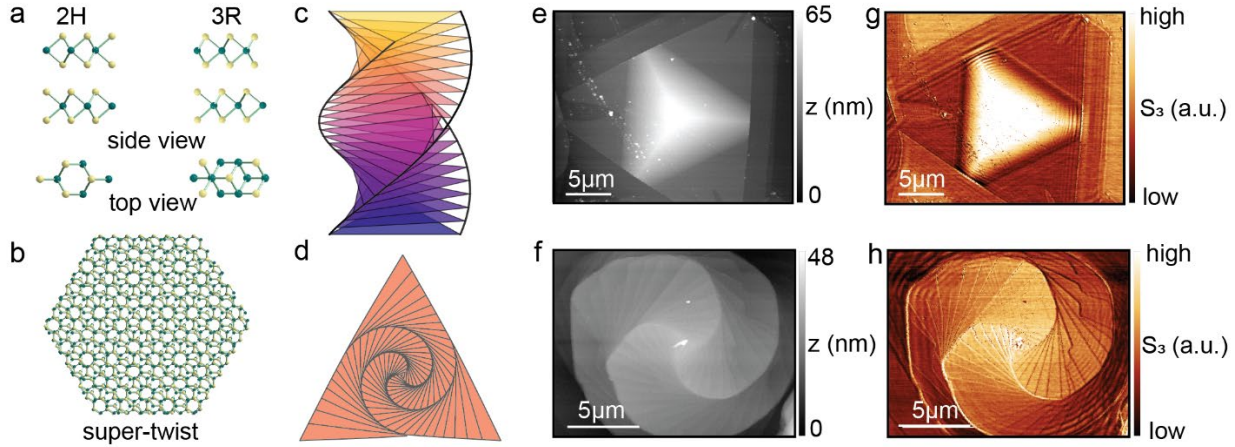

**Fig. 1. Structures of non-twisted and supertwisted spiral TMDs. a.** Side and top views of the most common 2H and 3R prototypes for multilayer TMD materials. The yellow and green balls represent chalcogen and transition metal atoms, respectively. **b.** Top view of a tetralayer WS₂ structure with an interlayer twist of 30°, showing the disruption of translational symmetry and the emergence of rotational symmetry. **c.** Side-view illustration of a 9°-supertwisted structure with exaggerated spacings and with each triangle representing a single layer of the TMD. **d.** Top-view sketch of a 9°-supertwisted spiral structure in which the lateral size of each successive layer decreases as the material grows from bottom to top. **e-f.** Atomic force microscopy images of representative non-twisted (e) and twisted (f) spirals of WS₂. **g-h.** Scanning near-field microscopy images of the same non-twisted (g) and twisted (h) WS₂ spirals as in e and f, respectively.

The supertwisted spiral structure can be classified under a nonsymmorphic screw symmetry, where a nonprimitive lattice translational vector along the z-axis is combined with rotational symmetry about the screw axis[30]. In contrast with other screw rotational symmetry materials [31, 32], the z-translation vector of a supertwisted TMD is not half or quarters of the lattice vector but is rather determined by the interlayer distance, $d_z$. The inverse of this distance, $1/d_z$, defines a quasi-momentum[24], $k_z$, which modulates the electronic structure by influencing interlayer hopping. With repeated Brillouin zone folding in both 2D and 3D, the electronic bands of supertwisted spiral TMDs become increasingly dense and further flattened by moiré hybridization. Both the in-plane Hamiltonian and interlayer tunneling are governed by the underlying rotational symmetry, cultivating emerging novel optical and electronic properties, as demonstrated in the following sections.

## Emergent excitons in supertwisted WS₂ spirals

We performed systematic PL measurements of the supertwisted WS₂ samples by varying the position (Fig. 2a), temperature (Fig. 2b), and supertwist angle (Extended Data 5). Fig. 2 displays the PL spectra of supertwisted $WS_2$ with an interlayer twist ($\alpha$) of 6°, taken at 10 K at six different positions from the center to the edge corresponding to different thicknesses. These measured positions are indicated by dark to light blue circles in the optical image (inset of Fig. 2a). Three distinct groups of PL features can be identified for the positions thicker than 10 nm (positions 1-4 of Fig. 2a). The center (position 1) measures around 40 nm in height, according to the atomic force microscope (AFM) (inset of Fig. 2b). Noticeably, the first group of PL peaks (labeled A in Fig. 2a-b), observed at around 610 nm, resemble the conventional A exciton[33,34] and its composites.[33-36] in $WS_2$ layers. Based on the order-of-energy comparison, it is reasonable to attribute the small splitting features at 602, 609 and 616 nm to A exciton, trion and biexciton states, respectively. The intensities of these group-A excitons first decrease and then increase from position 1 to position 6. While the $WS_2$ layers at positions 5 and 6 are thin (<10 nm), positions 1-4 are more bulk-like (10-40 nm) with PL features emerging at around 740 nm (labeled E) and 820 nm (labeled I) in the spectra. While the group-A excitons can be traced from conventional 2D $WS_2$ excitonic responses[32-37], the emerging second and third groups of peaks in Fig. 2a have no 2D counterparts. Furthermore, a prominent enhancement of the oscillator strength by the thickness can be observed for all peaks in positions 1-4. This enhancement can be roughly understood by considering the emission from an ensemble of uncoupled oscillators[38] with oscillator strength $f = \frac{2N\omega_0}{\varepsilon_0\hbar}\left|H_{ij}^D\right|^2$ , where $\omega_0$, $\varepsilon_0$, $\hbar$, $N$, and $H_{ij}^D$ are the frequency, dielectric constant, Planck constant, number of oscillators per unit volume, and optical dipole transition matrix, respectively. With increasing material thickness, the oscillator density $N$ increases and therefore the oscillator strength $f$ increases; this is similar to the observation in ref [39] at room temperature.  The temperature dependence of the PL spectra (Fig. 2b) provides further insight into the origins of the new excitons emerging at the thick central region of the supertwisted $WS_2$. The A exciton peaks and I peaks are observed at all temperatures, whereas the E peaks are only prominent below 160 K. The slight variation of the A exciton intensity below 160 K is potentially due to competition between 2D and 3D effects, which can also be observed in the thickness dependence (Fig. 2a). Intriguingly, the emerging E and I peaks exhibit opposite and monotonic temperature dependencies: lowering the temperature increases the intensity of the E peaks whereas it decreases the intensity of the I peaks. These behaviors indicate a direct transition nature for the former and a momentum-indirect transition for the latter. Defect-bound states are ruled out as the origin of these new excitonic features because they are absent in the non-twisted $WS_2$ samples from the same synthesis (Extended Data Fig. 4) and strongly depend on the supertwist angles (Extended Data Fig. 5).

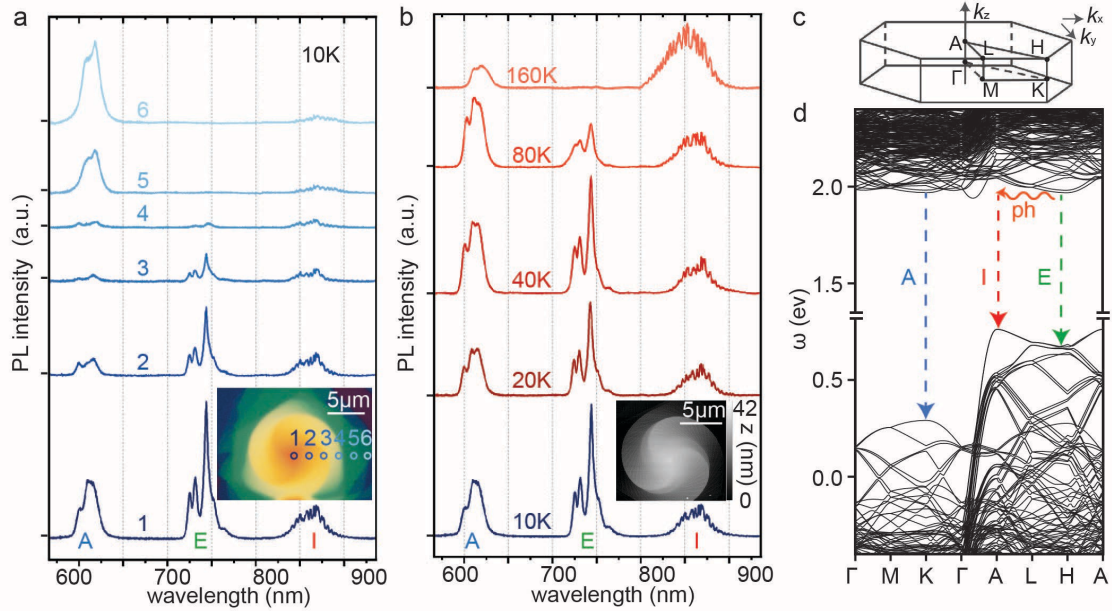

**Fig. 2. Photoluminescence (PL) spectra showing emerging exciton features in α = 6° supertwisted WS₂. a.** PL spectra at six different locations from center to edge (1-6, dark-light blue) of supertwisted WS₂ with α = 6° measured at T = 10 K. Inset: an optical image of the sample with measurement locations marked. The colored letters A, E, and I at the bottom mark the three main types of exciton features, which are later indicated by dashed arrows of the same color in panel d. **b.** PL spectra at the center (position 1 in a) at different temperatures from 160 K to 10 K. Inset: AFM topography image of the sample. **c.** 3D Brillouin zone definition for the band structure shown in d. **d.** Band structure of bulk supertwisted WS₂. The dashed blue arrow (A) marks the transitions around the K point corresponding to A excitons. The dashed green arrow (E) denotes the transitions around the H point corresponding to emerging excitons from the new bandgap due to the third-dimensional rotation. The red arrow (I) indicates the indirect excitonic transitions assisted by phonons marked by the wiggling orange arrow.

To understand the origin of these emergent bright excitons, we calculate the electronic band structure (Fig. 2d) using generalized Bloch theory[24], with screw rotational symmetry substituting for translational symmetry. Correlating the PL spectra with the band structure allows us to assign the three types of excitons to optical transitions. As indicated by the blue dashed arrow in Fig. 2d, the A excitons are excited around the K point, which is similar to what occurs in monolayer WS₂. The flattened band shape provides a large effective mass of A excitons in supertwisted WS₂, in contrast to the indirect gaps for non-twisted WS₂. It is important to note that the flat bands in the 3D Brillouin zone (A-L-H-A) support the emergence of E and I excitons, indicated by dashed green and red arrows, respectively. While the origin of the splittings observed in the 3D E excitons remains unclear, our experiments confirm that these peaks are very sensitive to the supertwist angle (Extended Data 5), and robust in energies when varying the thickness and temperature (Fig. 2). On the other hand, the fine splitting structure of the I (indirect transition) peaks can be well understood by the phonon cascade effect[40,41,42]. As analyzed in detail in Extended Data 6, the Fourier transform (FT) of the I emission displays distinct energy peaks comparable to the 3D phonons of breathing and shear modes[43,44,45]. This suggests that during the indirect exciton emission process, integer numbers of interlayer phonons are emitted or absorbed before the exciton recombines, providing momentum for the indirect transitions and forming the split emission spectra. These splittings are detectable with both charge-coupled device (CCD) and InGaAs detectors, ruling out optical etaloning artifacts associated with CCD measurements. The robustness of the A excitons, along with the distinct behaviors of the emerging E and I excitons, reflects the unique band properties inherited from the 3D supertwisted structure.

To gain a complete view of the three types of excitons, we further conducted wide-field imaging with polarimetry to obtain real-space distribution and polarization information for the A, E, and I exciton emissions. By exciting the sample with a linear 532-nm laser above the bandgaps, we were able to measure the intrinsic emission with a polarization analyzer and in wide field (Methods). A complete representation of the polarization state of electromagnetic radiation is typically described by a Stokes vector, $(S_0, S_1, S_2, S_3)$ in Cartesian coordinates or $(I, p, \chi, \Psi)$ in spherical coordinates, forming a Poincaré sphere (Fig. 3a). Unlike the single-point spectroscopy used in Fig. 2a-b, we record the exciton emission from the whole sample with a wide-field camera. Following the excitation, emissions from A, E, and I excitons were separated using bandpass filters and analyzed with an achromatic quarter-wave plate (QWP) and a linear polarizer. Stokes parameter maps for the A, E, and I excitons were obtained by rotating the QWP through 18 different angles (Supplemental Video 1-3) with standard analysis[46]. In Figs. 3b, c and d, the intensity parameters $(I = S_0)$ for the A, E, and I excitons are displayed in blue, green, and red, respectively. The A exciton exhibits a clear 2D character (Fig. 3b) and dominates the thinner outer regions of the sample. In contrast, the intensity maps of the E (Fig. 3c) and I (Fig. 3d) excitons display 3D characteristics, with stronger intensities localized in the thicker center and some edges. Overlaying the images of the three excitons (Fig. 3b-d) produces Fig. 3e, where the three exciton types together form a rich texture of emission. In Figs. 3f, g and h, we present the degree of polarization parameter p = $\frac{\sqrt{S_1^2 + S_2^2 + S_3^2}}{S_0}$ for A, E, and I excitons, respectively. The non-zero p values reveal the intrinsically polarized states of the emergent excitons, which carry spin and valley information for the electronic structures. As detailed in Extended Data 7, the indirect I excitons exhibit both more pronounced circular polarization ($\chi$) and linear polarization ($\Psi$) than the direct E excitons, potentially due to the spatial separation of dipole moments[47] and the involvement of phonon scattering[48]. Although the supertwisted structure complicates the establishment of optical selection rules and spin-orbit coupling, the maps of the PL intensity and polarization states highlight the real-space distribution of different species of flat band excitons and their polarized nature.

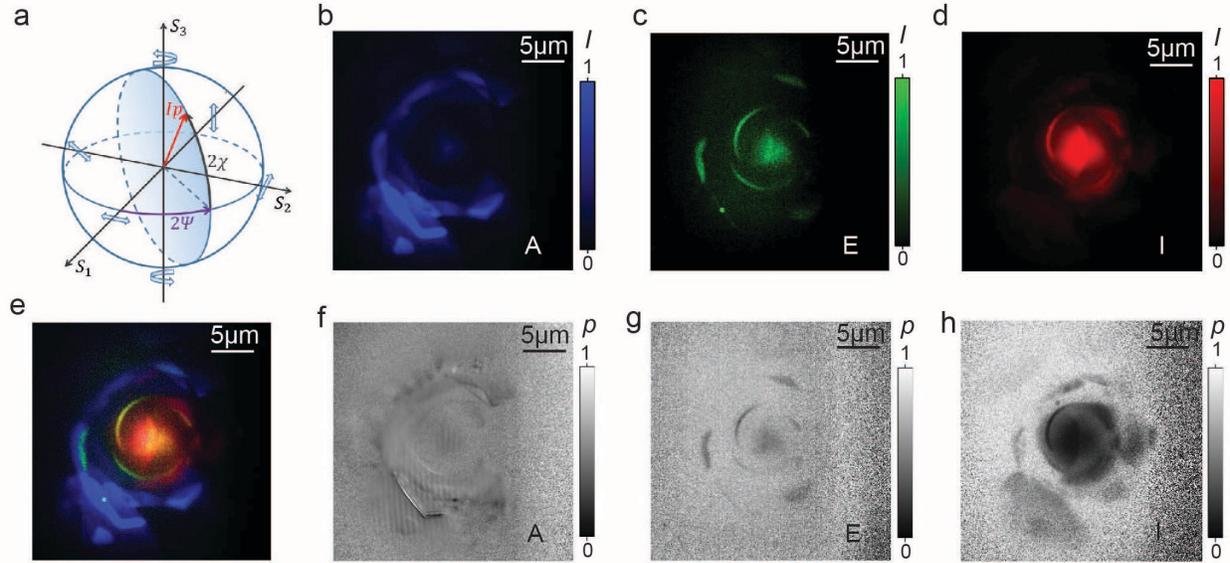

**Fig. 3. Wide-field polarimetry imaging of A, E, and I exciton emissions of a supertwisted WS$_2$ sample.** **a.** Standard Poincaré sphere coordinates for the Stokes representation of the polarization state of light. The Stokes vector can be expressed in Cartesian coordinates as (S$_0$, S$_1$, S$_2$, S$_3$) or in spherical coordinates as (I, p, $\chi$, $\Psi$). **b-d.** Intensity element maps (I or $S_0$ parameter) for A, E, and I excitons, respectively. The false color is used to differentiate the energy difference of A, E, and I exciton emission. **e.** Total exciton emission intensity obtained by adding the energy intensities in b, c and d. **f-h.** Degree of polarization (p = $\frac{\sqrt{S_1^2+S_2^2+S_3^2}}{S_0}$) mapping for A, E, and I excitons, respectively.

## Local density of states at band edges

To elucidate the local electronic structure and its relationship to exciton features, we investigated the local density of states (LDOS) at band edges for different supertwist angles. Because the generalized Bloch theory[24] for band structure calculations in Fig. 2d becomes imprecise for larger supertwist angles, we further developed a real-space tight-binding model parametrized from first-principles density-functional theory[49] to capture the electronic structure of supertwisted WS$_2$ with arbitrary supertwist angle $\alpha$ (Fig. 4). We keep in-plane direction with open boundaries and apply periodic boundary conditions in the z direction so that the layer-resolved density of states is the same in each layer, corrected by a uniform rotation of $\alpha$.

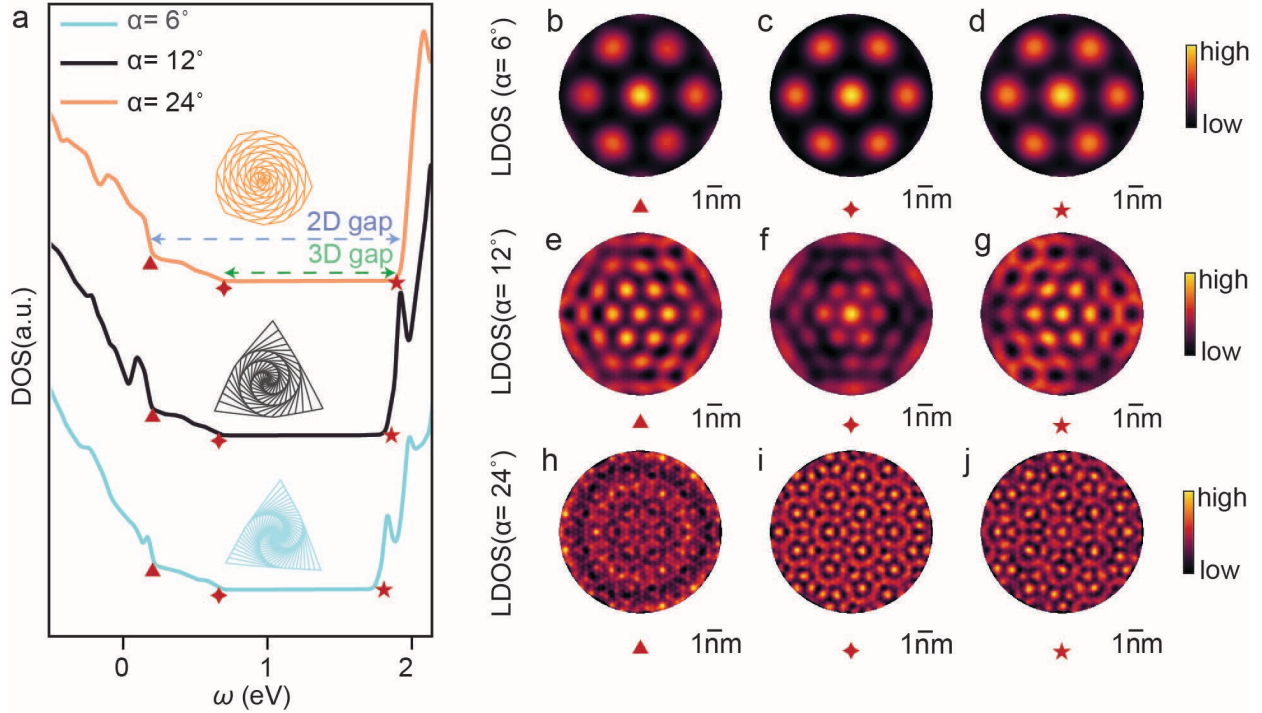

**Fig. 4. Real-space modeling of local density of states (LDOS) of supertwisted WS₂. a.** Total density of states of bulk supertwisted WS₂ with α = 6°, 12°, and 24°. The insets show sketches of corresponding supertwisted structures. The double-headed blue and green dashed arrows mark the 2D and 3D bandgaps, respectively. The stars mark the energies where 2D LDOS maps are illustrated. **b-d.** LDOS colormaps at the 2D and 3D band edge energies (marked by corresponding shapes shown in a) for α = 6°. **e-g.** LDOS colormaps at band edge energies for α = 12°. **h-j.** LDOS colormaps at band edge energies for α = 24°.

Figure 4a demonstrates representative results of the total density of states (DOS) for supertwisted WS₂ with α = 6°, 12°and 24° in blue, black and orange colors, respectively. The sketches of the corresponding supertwist structures are displayed above the DOS curves. Notably, the coexisting 2D and 3D energy gaps (denoted by the blue and green dashed arrows, respectively) are consistent with the observations of the band structure in Fig. 2d from the generalized Bloch theory[24]. We then analyze the origins of the flat band excitons from a real-space perspective. The colormaps in Fig. 4b-d demonstrate the real-space distribution of LDOS intensity at the 2D and 3D band edges of a supertwisted spiral of WS₂ with α = 6°; triangles, diamonds and stars mark the conduction band minimum (CBM) and 2D and 3D valence band maxima (VBMs), respectively. The strongly localized states in moiré periods (hotspots in Fig. 4b-d) corroborate the flat band observations in Fig. 2d. Interestingly, for α = 6°, the LDOS maps at the CBM and the 2D and 3D VBMs all align with each other in real space. This indicates that the wavefunctions for electrons and holes that form the direct A and E excitons strongly overlap with each other. The overlapped wavefunctions can lead to a large expectation value of the optical transition matrix element: $\left|H_{eh}^{D}\right|^{2} = |\langle\psi_{h}|H^{D}|\psi_{e}\rangle|^{2}$, where $\psi_{e}$ and $\psi_{h}$ are the initial and final states and $H^{D}$ is the dipole

operator. Therefore, the overall exciton oscillator strength, $f = \frac{2N\omega_0}{\varepsilon_0 \hbar}\left|H_{eh}^D\right|^2$, is enhanced and generates the bright PL seen in Figs. 2 and 3. The observation of overlapped hotspots at band edges is reminiscent of the exciton localization[50,51,52] in 2D moiré systems. Note that the LDOS hotspots are generally not aligned at energies deeper in the band, as demonstrated in Supplemental Videos 4-6. Instead, they might occasionally be compensating for each other, suggesting the formation of bonding-antibonding states. The LDOS maps at band edges for α = 12° (Fig. 4e-g) show that in addition to the overlapped LDOS hotspot positions, super-moiré (moiré of moiré) modulation appears and varies for different band edge energies. Moreover, for the largest angle α = 24°, the LDOS maps at band edges (Fig. 4f-j) no longer align with each other, suggesting that bright 3D exciton emission is not a universal phenomenon for arbitrary twist angles. This conforms to the angle-dependent PL in Extended Data 5. Although the super-moiré modulation seems to be more prominent for larger twist angles, the wavefunctions for band edges overlap better for α = 6°, underscoring the real-space origins of the observed bright flat band excitons.

## Summary and Outlook

Programmable 3D twistronics offers rich opportunities for engineering the electronic, photonic, and phononic properties of materials. Here, we reveal for the first time that a bulk supertwisted structure-semiconducting supertwisted WS₂-hosts bright flat band excitons. Through a combination of photoluminescence measurements and band structure calculations, we identify both 2D and 3D bright excitons arising from direct flat band transitions, as well as 3D indirect excitons mediated by phonon cascades. All emission features are enhanced with increasing thickness, approaching the 3D bulk limit. Wide-field PL polarimetry imaging further distinguishes the spatial distributions of the three exciton types and reveals that the 3D indirect excitons exhibit the strongest intrinsic polarization. Finally, the local density of states (LDOS) calculation illustrates the moiré localization and super moiré effects, whereby overlapped wavefunctions at band edges provide insights into the emergence of flat band excitons.

Although direct-gap transitions involving dipole-allowed processes explain much of the observed behavior, they do not tell the whole story. The presence of flat bands implies quenched carrier kinetic energy and enhanced Coulomb interactions, hosting correlation effects that are worthy of further investigation. While the flat band enhanced A excitons are sensitive to in-plane electronic transport, the E and I excitons may provide new sensitive probes of collective interlayer interactions[53] due to their 3D nature. Looking ahead, incorporating additional tuning parameters—such as structured light excitation, doping, and electric and magnetic fields—will lead to a greater understanding of bulk flat band properties, including the spiral nature of the electron wavefunctions[54, 55], the quantum geometry of flat band excitons[56], and the competition between 2D and 3D excitonic phenomena. The tunable flat band exciton emissions also offer opportunities for the development of quantum emitters[57] and further advances in chiral phonon dynamics[58] and topological polaritonics[59].

## Acknowledgements

This work was mainly supported by Programmable Quantum Materials, an Energy Frontier Research Center funded by the U.S. Department of Energy (DOE), Office of Science, Basic Energy Sciences (BES), under award no. DE-SC0019443. D.N.B. is a Moore Investigator in Quantum Materials EPIQS no. 9455. Y.Z. and S.J. are supported by the National Science Foundation DMREF program (NSF-2323470). L.K. was supported by the Excellence Initiative of the German federal and state governments and the Deutsche Forschungsgemeinschaft (DFG, German Research Foundation). L.K. and T.O.W. also gratefully acknowledge support from the DFG through FOR 5249 (QUAST, Project No. 449872909) and SPP 2244 (Project No. 422707584). EJM is supported by the Department of Energy under grant DE FG02 84ER45118. T.O.W. is supported by the Cluster of Excellence "CUI: Advanced Imaging of Matter" of the DFG (EXC 2056, Project ID 390715994). K.W.C.K. acknowledges support from the Department of Energy (DOE) National Nuclear Security Administration (NNSA) Laboratory Residency Graduate Fellowship program no. DE-NA0003960.

## Author contributions

Y.D. led the experiments and data analysis. Y.Z. and S.J. synthesized the materials. L.K. performed band structure and local density of states calculations, advised by T.W.. T.H. and Y.D. performed the single-point PL experiments. D.X. performed the wide-field PL experiments with help from Y.D. and J.A.T.. C.T. performed the second harmonic generation experiments with participation from T.P.D., K.W.C.K., and Y.B.. C.H. participated in angle-dependent PL analysis. D.S. and Y.D. performed Raman characterization. Y.D., Y.S. and M.F. performed near-field experiments. R.Q., M.D., P.J.S., X.Z., S.J., E.M., and T.W. advised the project and discussed the results. D.N.B. supervised the project.

## Methods

### Scanning near-field optical microscopy

The scanning near-field optical images were taken with NeaSNOM (Neaspec GmbH, Attocube) under an excitation with a continuous-wave 720-nm laser (M-squared Lasers Ltd). The tapping AFM tip above the sample demodulated the local scattered signal, with another reference dithering mirror demodulating the background. Using the pseudoheterodyne method, the near-field signals were extracted; they are background-free for higher-order demodulated signals. The resolution of the near-field images was 15 nm, as determined by the apex size of the AFM probe.

The waveguide modes propagating inside the crystal could be imaged by means of the SNOM experiments.

## Second harmonic generation mapping

A tunable optical parametric oscillator (OPO) laser with an 80MHz repetition rate was used to pump the second harmonic generation (SHG) of the supertwisted $WS_2$ spirals. The pump laser wavelength was set to 1240 nm, 1280 nm, 1300 nm, 1350 nm and 1500 nm to generate SHG signals across the 2D and 3D direct exciton energy ranges. The polarization of the pump was controlled by a rotating broadband half-wave plate (1100-2000 nm). The pump laser was focused on the sample using a 100x NA=0.95 objective with pump power fixed at 1.5 mW. The sample was located on a motorized piezo stage for spatial positioning and SHG mapping. The emitted SHG was collected in a reflection geometry by the same objective, with the analyzer polarizer kept parallel to the pump polarization. Finally, the SHG signal was detected by an EMCCD.

## Spatially and temperature-dependent photoluminescence spectroscopy

The spatially and temperature-dependent PL spectra were measured under an excitation with a continuous-wave 532-nm laser (Lighthouse Photonics). The excitation power was 250 µW unless otherwise specified. The excitation laser was first sent through a bandpass filter and then a 100x objective lens equipped closed-cycle He cryostat (Montana Instruments). The sample position was controlled by piezoelectric motors (Attocube). The PL from the sample was collected using the same objective, sent through long-pass filters to block the excitation laser, and then sent to a spectrometer where a monochromator dispersed the PL onto a liquid $N_2$-cooled charge-coupled device or InGaAs array detector for detection.

## Wide-field photoluminescence imaging and polarimetry

The wide-field PL imaging was also measured under an excitation with a continuous-wave 532-nm laser, with the sample placed in a Montana He cryostat with a 100x objective lens. An f = 250-mm wide-field lens focused the excitation laser beam in the back focal plane of the objective and generated wide-field illumination of the sample. The PL from the sample was then spatially imaged on a CMOS camera (Kiralux, Thorlabs) after passing through the imaging lenses. An achromatic quarter-wave plate (690-1100 nm) and a linear polarizer (650-1100 nm) were inserted in front of the last imaging lens for measuring the Stokes polarization parameters. A color filter was placed before the camera to select A-, I-, or E-exciton emission individually. We used the rotating quarter-wave plate measurement explained in ref [36] to measure the Stokes parameters.

The Stokes parameters were measured by monitoring the evolving emission intensity when rotating a quarter-wave plate before a linear polarizer. The quarter-wave plate creates a relative phase shift between the orthogonal components of the supertwisted $WS_2$ emission. We acquired three series of angle-resolved PL images of A-, I-, and E-exciton emission as $I_n^A(\vec{r}, \theta_n)$, $I_n^I(\vec{r}, \theta_n)$, and $I_n^E(\vec{r}, \theta_n)$ with different edge-pass color filters. Each series consists of 18 images on the quarter-wave plate at angles ranging from 0 to 170 degrees with 10-degree spacing. Data processing follows the method in ref [36]. Due to the spatial inhomogeneity of the twisted $WS_2$, the Stokes parameters vary across different regions of the crystal. Consequently, we analyzed the angle-resolved PL images to generate a map of the Stokes parameters. From our rotating polarizing optics analyzer system, we see that the intensity of the twisted $WS_2$ emission on each detector pixel is given by:

$$I\ (\vec{r},\ \theta) = \frac{1}{2}[\ S_0(\vec{r}) + S_1(\vec{r})\cos^2 2\theta + S_2(\vec{r})\cos 2\theta\sin 2\theta + S_3(\vec{r})\sin 2\theta\ ] \tag{1}$$

where $I\ (\vec{r},\ \theta)$ is the spatial emission intensity at a given angle θ and $S_n(\vec{r})$, n=1-3 are the spatially resolved Stokes vectors of the twisted $WS_2$ crystal. The Stokes parameter map data are extracted with the conversion matrix as follows:

$$\begin{pmatrix} S_0(\vec{r}) \\ S_1(\vec{r}) \\ S_2(\vec{r}) \\ S_3(\vec{r}) \end{pmatrix} = \frac{4}{n}\begin{pmatrix} 1 & 0 & -1 & 0 \\ 0 & 0 & 2 & 0 \\ 0 & 0 & 0 & 2 \\ 0 & 1 & 0 & 0 \end{pmatrix}\begin{pmatrix} 1/2\sum I_n \\ \sum I_n\sin 2\theta_n \\ \sum I_n\cos 4\theta_n \\ \sum I_n\sin 4\theta_n \end{pmatrix} \tag{2}$$

The Stokes parameter maps are presented as the standard Poincaré sphere coordinates $I$, $p$, $\psi$, and $\chi$, as shown in Extended Data Fig. 6. The conversion relation is:

$$\begin{cases} I = S_0 \\ p = \sqrt{S_1^2 + S_2^2 + S_3^2}/S_0 \\ 2\psi = \arctan\frac{S_2}{S_1} \\ 2\chi = \arctan\frac{S_3}{\sqrt{S_1^2 + S_2^2}} \end{cases} \tag{3}$$

# Supporting Information for

## Flat band excitons in a three-dimensional supertwisted spiral transition metal dichalcogenide

**Yinan Dong, Yuzhou Zhao, Lennart Klebl, Taketo Handa, Ding Xu, Chiara Trovatello, Chennan He, Dihao Sun, Thomas P. Darlington, Kevin W. C. Kwock, Jakhangirkhodja A. Tulyagankhodjaev, Yusong Bai, Yinming Shao, Matthew Fu, Raquel Queiroz, Milan Delor, P. James Schuck, Xiaoyang Zhu, Tim Wehling, Song Jin, Eugene Mele, Dmitri N. Basov**

**Yinan Dong, Dmitri N. Basov**
**E-mail: yd2400@columbia.edu, db3056@columbia.edu**

**This PDF file includes:**

**Other supporting materials for this manuscript include the following:**

## Supporting Information Text

## 1. Tight-binding Model

We model the electronic structure of WS$_2$ spirals in the tight-binding approximation. We start with a description of the monolayer Hamiltonian in Section A and thereafter continue with the interlayer hoppings in Section B. The code that implements the tight-binding model as well as the real-space calculations ( Section 3) is publicly available under Ref. (1).

**A. Monolayer WS$_2$.** The electronic structure of WS$_2$ monolayers is well described by an eleven-orbital model (2), where the W-$d$ shell is taken into account as well as the $s-p$ shell. The unit cell is triangular with lattice vectors $\boldsymbol{a}_1 = a_0 \, (1,0,0)^T$ and $\boldsymbol{a}_2 = a_0 \, (\cos 2\pi/3, \sin 2\pi/3, 0)^T$, where $a_0 = 3.1532 \, \text{Å}$. The tungsten atoms occupy the honeycomb-$A$ sites and the sulfur atoms are located at the honeycomb-$B$ sites shifted up and down symmetrically in $z$-direction, i.e.,

$$\boldsymbol{b}_{\text{W}} = \frac{2\boldsymbol{a}_1 + \boldsymbol{a}_2}{3}, \qquad \boldsymbol{b}_{\text{S}_{12}} = \frac{\boldsymbol{a}_1 + 2\boldsymbol{a}_2}{3} \pm |\boldsymbol{b}_{\text{W}}|(0,0,\tan\theta) \,, \tag{1}$$

where we use $\theta = 0.710$ (radiant). The monolayer tight-binding Hamiltonian is set up in real-space as follows:

$$H^M = \sum_{\boldsymbol{R}_i, \boldsymbol{R}_j, o, p, \sigma} c^\dagger_{\boldsymbol{R}_i, o, \sigma} V(\boldsymbol{R}_j - \boldsymbol{R}_i, o, p) c_{\boldsymbol{R}_j, p, \sigma} + H^{\text{SOC}} \,, \tag{2}$$

with $c^{(\dagger)}_{\boldsymbol{R}_i, o, \sigma}$ annihilating (creating) an electron with spin $\sigma$ in orbital $o$ and unit cell $\boldsymbol{R}_i$. The hopping parameters $V(\boldsymbol{R}, o, p)$ follow the Slater-Koster (SK) parametrization from Ref. (3). We also use the atomic $\boldsymbol{L} \cdot \boldsymbol{S}$ spin-orbit coupling parametrization of Ref. (3) that is abbreviated as $H^{\text{SOC}}$.

**B. Interlayer hoppings.** For any twisted heterostructure, we require a parametrization of interlayer hoppings for arbitrary distances and orientations of sites. The (twisted) $N$-layer Hamiltonian takes the following form:

$$H = \sum_{l=1}^{N} H_l^M + \sum_{\sigma} \left[ \sum_{l=1}^{N-1} \sum_{\boldsymbol{R}_i^l, o^l, \boldsymbol{R}_j^{l+1}, p^{l+1}} c^\dagger_{\boldsymbol{R}_i^l, o^l, \sigma} \mathcal{V}(\boldsymbol{R}_j^{l+1} - \boldsymbol{R}_i^l, o^l, p^{l+1}) c_{\boldsymbol{R}_j^{l+1}, p^{l+1}, \sigma} + \text{h.c.} \right] \equiv \sum_{l=1}^{N} H_l^M + \sum_{l=1}^{N-1} \left[ H_l^I + \text{h.c.} \right], \tag{3}$$

where $H_l^M$ is the implementation of Eq. (2) for each individual layer and $H_l^I$ denotes the interlayer Hamiltonian. The interlayer SK hopping parameters $\mathcal{V}(\boldsymbol{R}, o, p)$ are taken from Ref. (4). Notably, the lattice may differ from layer to layer ($\boldsymbol{R}_i^l$ and $o^l$ carry a layer index $l$), which is how we enable the description of twisted structures. Geometrically, we separate each WS$_2$ layer by a distance of $c_0 = 5.404 \, \text{Å}$. We assume the layers to be rigidly stacked, as there is no single moiré interface that would allow for trivial out-of-plane relaxation due to the spiral nature of the samples.

## 2. Small-angle Approximation

For small twist angles, we make use of the construction presented in Ref. (5), which allows to obtain the band structure of spirally twisted systems through exploitation of the nonsymmorphic layer translation symmetry: Assuming an infinite layer number or periodic boundary conditions and a relative twist angle of $\theta$ in between each of the layers, the Hamiltonian Eq. (3) is invariant under screw symmetry $\mathcal{S} = \mathcal{R}_{-\theta}\mathcal{T}_z$:

$$\mathcal{S}H = \mathcal{R}_{-\theta}\mathcal{T}_z H = \mathcal{R}_{-\theta}\left( \sum_l H_{l+1}^M + \sum_l \left[ H_{l+1}^I + \text{h.c.} \right] \right) = \sum_l \mathcal{R}_{-\theta}H_{l+1}^M + \sum_l \left[ \mathcal{R}_{-\theta}H_{l+1}^I + \text{h.c.} \right]$$

$$= \sum_l H_l^M + \sum_l \left[ H_l^I + \text{h.c.} \right] = H \,, \quad [4]$$

with layer translation $\mathcal{T}_z$ and $\mathcal{R}_\theta$ the rotation by an angle of $\theta$. One can make use of the screw symmetry to formulate the generalized Bloch theorem. When transforming to screw momentum space in $z$ direction, i.e.,

$$c_{\boldsymbol{r}, k_z} = \frac{1}{\sqrt{N}} \sum_l e^{-ilk_z} c_{\mathcal{R}_{n\theta}\boldsymbol{r}, l} \,, \tag{5}$$

and in small-angle approximation, the Hamiltonian becomes (5)

$$H = \sum_{k_z} H^M + H^I e^{ik_z} + (H^I)^\dagger e^{-ik_z} \,. \tag{6}$$

In the practical implementation of Eq. (6) starting from the SK model (cf. Sections A and B), one has to define a commensurate moiré supercell. Then, the monolayer Hamiltonian is defined on discrete atom positions of a single (the bottom) layer of the moiré cell, whereas the interlayer Hamiltonian is defined on *different* positions, i.e., the positions of the bottom *and* the top layers. In order to mimic the continuum approximation of Ref. (5), we use the intralayer and interlayer Hamiltonians from the commensurate moiré bilayer and find the closest bottom-layer position (with the right orbital) for each top-layer position. That way, we can map the interlayer Hamiltonian to the discrete real-space of a single layer and solve Eq. (6) in the tight-binding basis; going to the two-dimensional moiré Brillouin zone and using $k_z$ as third momentum.

## 3. Real-space Model

Since the experimental structures are not firmly placed in the small-angle regime, we perform real-space calculations that are independent on the twist angle. We follow the atomic structure and Hamiltonian setup specified in Sections A and B, except for the fact that we now have to truncate the number of layers $N_l$, as well as the two-dimensional extent of the overall flake (there is no commensurate moiré cell as we twist in the same direction with each subsequent layer). We use a circular cutoff (in order to not break any symmetries), i.e., only positions with $|\boldsymbol{r}|_{2\mathrm{D}} < R_c$ are kept in the flake. Depending on $N_l$ and $R_c$, the Hamiltonian size is up to roughly $10^7$, so we save it as a sparse matrix. We do not diagonalize the Hamiltonian, but instead analyze spectral properties by means of the kernel polynomial method (KPM) (6, 7).

**A. Kernel polynomial method.** In order to make use of KPM, one performs a linear transformation on the Hamiltonian $H \rightarrow \tilde{H}$ that maps its spectrum to the window $(-1, 1)$. For this system, we know that the transformation

$$\tilde{H} = (H - b)/a \,, \qquad a = \frac{E_{\max} - E_{\min}}{2} \,, \qquad b = \frac{E_{\max} + E_{\min}}{2} \,, \qquad E_{\min} = -15\,\mathrm{eV} \,, \qquad E_{\max} = 5.5\,\mathrm{eV} \qquad [7]$$

places sufficient bounds on the spectrum $\sigma(H)$ such that $\sigma(\tilde{H}) \in (-1, 1)$. As detailed in Ref. (6), the (local) density of states of $\tilde{H}$ (DOS, LDOS) can be expressed by

$$f(\tilde{\omega}) = \frac{1}{\pi\sqrt{1 - \tilde{\omega}^2}} \left[ \mu_0 + 2\sum_{n=1}^{\infty} \mu_n T_n(\tilde{\omega}) \right] \,, \qquad [8]$$

with

$$\mu_n = \frac{1}{D}\mathrm{Tr}\big(T_n(\tilde{H})\big) \text{ for the DOS } \rho(\tilde{\omega}) \text{ as } f(\tilde{\omega}) \,, \qquad \mu_n = \frac{1}{D}\langle i|T_n(\tilde{H})|i\rangle \text{ for the LDOS } \rho_i(\tilde{\omega}) \text{ as } f(\tilde{\omega}) \,. \qquad [9]$$

In practice, the moments $\mu_n$ can be evaluated recursively, with either a stochastic evaluation of the trace or usage of a reference state $|i\rangle$ (in case of the LDOS). $D$ denotes the dimension of $\tilde{H}$. In any case, the expansion must be truncated; we denote the maximum order of Chebyshev polynomial by $M$ and use $M = 1500$ throughout. For random sampling, one makes use of the fact that

$$\mathrm{Tr}\big(T_n(\tilde{H})\big) \approx \frac{1}{R}\sum_{r=1}^{R}\langle r|T_n(\tilde{H})|r\rangle \,, \qquad [10]$$

with $|r\rangle$ random states (weights on each orbital randomly distributed on the complex unit circle). We use $R = 10$, which is sufficient at large system sizes.

To prevent oscillatory behavior of the (L)DOS, we use the "Jackson Kernel" (6) combined with the "Chebyshev polynomial Green's function" kernel (7), which amounts to replacing the moments $\mu_n \rightarrow \tilde{\mu}_n$, where

$$\tilde{\mu}_n = \mu_n g_n^J g_n^G \,, \quad g_n^J = \frac{1}{N+1}\left[(M - n + 1)\cos\frac{\pi n}{M+1} + \sin\frac{\pi n}{M+1}\cot\frac{\pi}{M+1}\right] \,, \quad g_n^G = \frac{2ie^{in\arccos(\tilde{\omega} - i\tilde{\eta})}}{\sqrt{1 - (\tilde{\omega} - i\tilde{\eta})^2}} \,. \qquad [11]$$

The broadening $\tilde{\eta} = \eta/a$ is set to $\eta = 10^{-3}\,\mathrm{eV}$.

**B. Geometry.** As parameters for the system's geometry, we use $N_l = 20$ layers whenever considering the real-space model. The angles are fixed at $\theta \in \{6°, 12°, 24°\}$. For the DOS calculations, we use a cutoff radius of $R_c = 90|\boldsymbol{a}_1|$. We additionally enforce the randomly sampled vectors $|r\rangle$ to be nonzero only in a range such that $|\langle r|\hat{\boldsymbol{r}}|r\rangle|_{2\mathrm{D}} \leq R_c - R_d$ (with $R_d = 5|\boldsymbol{a}_1|$). This ensures that the sampled DOS contains only bulk effects, effectively turning off the boundary.

As the LDOS calculations are numerically much more challenging than sampling the DOS (we cannot use statistical trace evaluation), we resort to slightly smaller system sizes with $R_c = 50|\boldsymbol{a}_1|$. The sites/orbitals that are sampled are taken from the top layer and satisfy $|\langle i|\hat{\boldsymbol{r}}|i\rangle|_{2\mathrm{D}} \leq R_l$ with $R_l = 15|\boldsymbol{a}_1|$.

## 4. Nonlinear Optical Properties

The supertwisted spiral TMDs are generally noncentrosymmetric and are thus expected to exhibit strong nonlinear optical effects. We performed second-harmonic generation (SHG, Figs. S1–S2) and Raman spectroscopy (Fig. S3) to characterize the nonlinear optical consequences of the supertwist.

**A. SHG.** The SHG intensities from bulk regions can be either enhanced or suppressed compared to few-layer, non-twisted base area, as shown in Fig. S1. The layer-by-layer supertwist significantly alters the quasi-phase-matching conditions, leading to rich spatial distributions of SHG signals which are worth further investigation. In Fig. S2, we focused on the main sample with $\alpha = 6°$ for position-, energy-, and polarization-dependent SHG measurements around the flatband excitons. A six-fold petal-shaped pattern characteristic of $C_3$ rotational symmetry was preserved in the absence of strain (8), as shown in Figs. S2a-e. Remarkably, the energy dependence, rather than strain, appears to stretch the petals. The summary of the SHG polarization in Fig. S2f and i (i is the overall shift of f) clearly reveal the excitation energy dependence of the SHG polarization, in addition to the position dependence with supertwist effect. Therefore, the SHG polarization is primarily determined by the orientation of the top layer and further modulated by the interference with the electronic wavefunctions in underlying layers. The interplay between structural supertwist, light wavelength effect and electronic properties together contribute to the rich spatial distribution of SHG amplitudes across different energies (Figs. S2g-k).

Notably, when the SHG energies are close to be resonant with 2D and 3D flatband excitons ($\lambda_{2\omega} = 620nm, 750nm$, respectively), both enhanced absorption and significant modifications in polarization distribution are observed. For the 2D exciton resonance case (pink curve in Fig. S2f), the SHG polarization becomes more broadly distributed across different spatial regions compared to off-resonant conditions. In contrast, when $2\omega$ is resonant with the 3D E exciton (purple curve), the spatial variation in polarization is slightly reduced. This can be understood by the electronic polarization is strongly altered by excitons of different ith different excitation energy

**B. Raman.** The two Raman active optical phonon modes, $E_{2g}$ and $A_{1g}$, are both blue-shifted when moving from the thinner sample edge to the thicker center, conforming to regular thickness-induced phonon stiffening behavior. The periodic modulation of their intensities is caused by a combination of the layer rotation and polarization dependence. Low-frequency Raman, which is not accessible by the time this manuscript is written, could be a particularly interesting future direction with the phonon cascade effect revealed in Fig. S6.

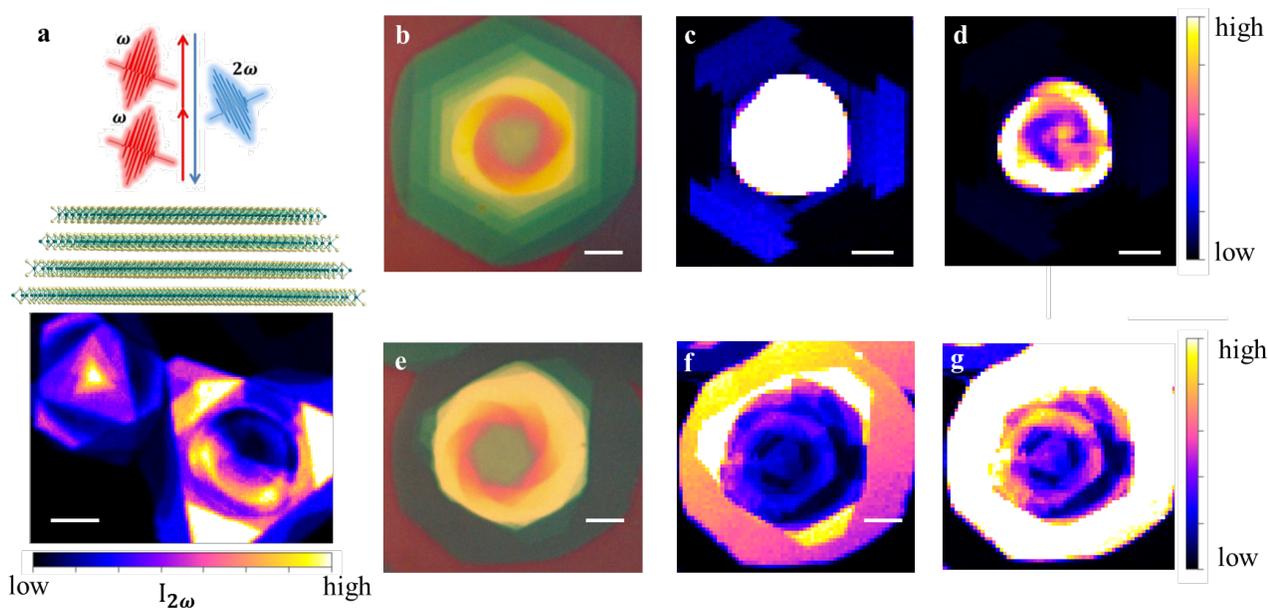

**Fig. S1.** Diverse second harmonic generation (SHG) responses in supertwisted spiral WS$_2$. a, Schematic of SHG experiment. The lower part in a shows a comparison of uniform (top left) and non-uniform (bottom right) supertwisted spiral WS$_2$. For the latter, an uneven base layer affects the overall SHG response. b and e, optical pictures of two prototype super-twisted spirals of three-fold and six-fold macroscopic rotational structures. The SHG signals at center areas can be either enhanced or decreased resembling R- or H- stacking behaviors. c and d, SHG intensity maps of flake in b. f and g, SHG intensity maps of flake in e. c and f are of the same color scale while d and g are of the same color scale zoomed from c and f. Scale bars in all images, 5 $\mu$m. Excitation laser wavelength in all SHG maps, 800nm.

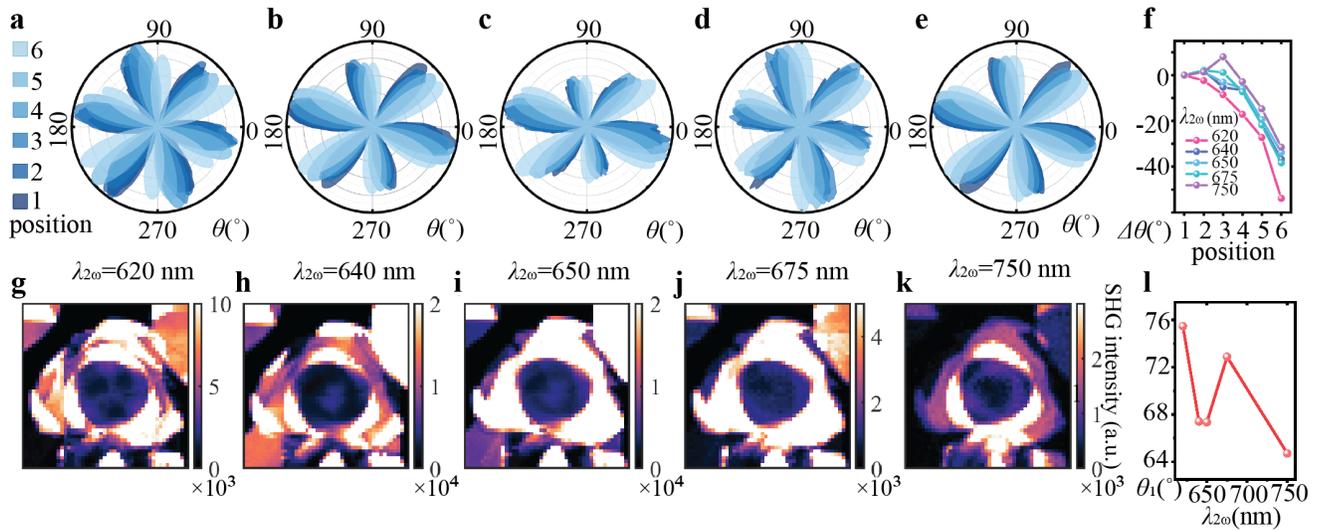

**Fig. S2.** Position-, energy-, and polarization-dependent SHG on the main sample of supertwist angle $\alpha = 6°$. a-e, Polarization-dependent SHG at the same six positions from center to edge (1-6) of the same sample as Fig. 2-3. The six polar map panels in a-e are corresponding to six excitation laser energy with the second-harmonic energy $\lambda_{2\omega} = 620\,nm, 640\,nm, 650\,nm, 675\,nm, 750\,nm$ as labeled underneath the polar maps. The six-fold polarization dependence reveals that the $C_3$ rotational symmetry is preserved for supertwist WS$_2$. The energy dependent stretch of six-fold pedals defies the lattice strain origin. f. Fitted angles ($\theta$) of the polar maps in a-e. The five curves are shifted to position 1 with offsets displayed in l. When $\lambda_{2\omega} = 620\,nm$ is close to resonance with 2D excitons, and when $\lambda_{2\omega} = 750\,nm$ is close to resonance with 3D excitons, $\Delta\theta$ shows abnormal distribution off the other energies' curves. The back-bending at position 3 for $\lambda_{2\omega} = 750\,nm$ is reproducible. g-k, Energy dependent SHG intensity maps with varied two-photon energy $\lambda_{2\omega} = 620\,nm, 640\,nm, 650\,nm, 675\,nm, 750\,nm$. The three-fold rotational textures vary with the different excitation energies. l. Energy dependent polarization angles ($\theta$) at the center of the material which is the offset taken at position 1 of f.

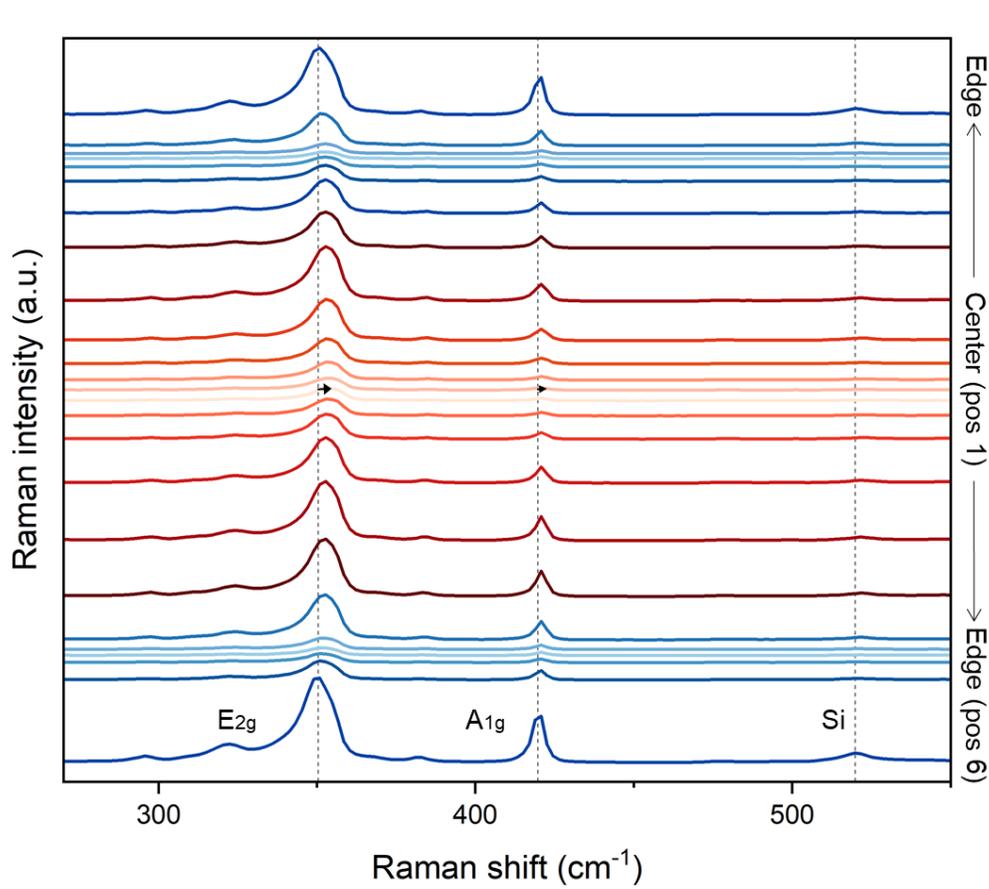

**Fig. S3.** Raman spectra on the sample of supertwist angle $\alpha = 6°$. The Raman spectra were taken across the spiral from edge to edge across the center with the corresponding locations marked on the right. The three vertical dashed lines indicate the E2g, A1g phonons of super-twisted WS2, and the TO phonon from Si substrate. From the thinner edge to the thicker center, both E2g and A1g phonons stiffen and therefore blue shifted. The periodic modulation of Raman intensity is caused by the layer rotation and polarization dependence of Raman intensity.

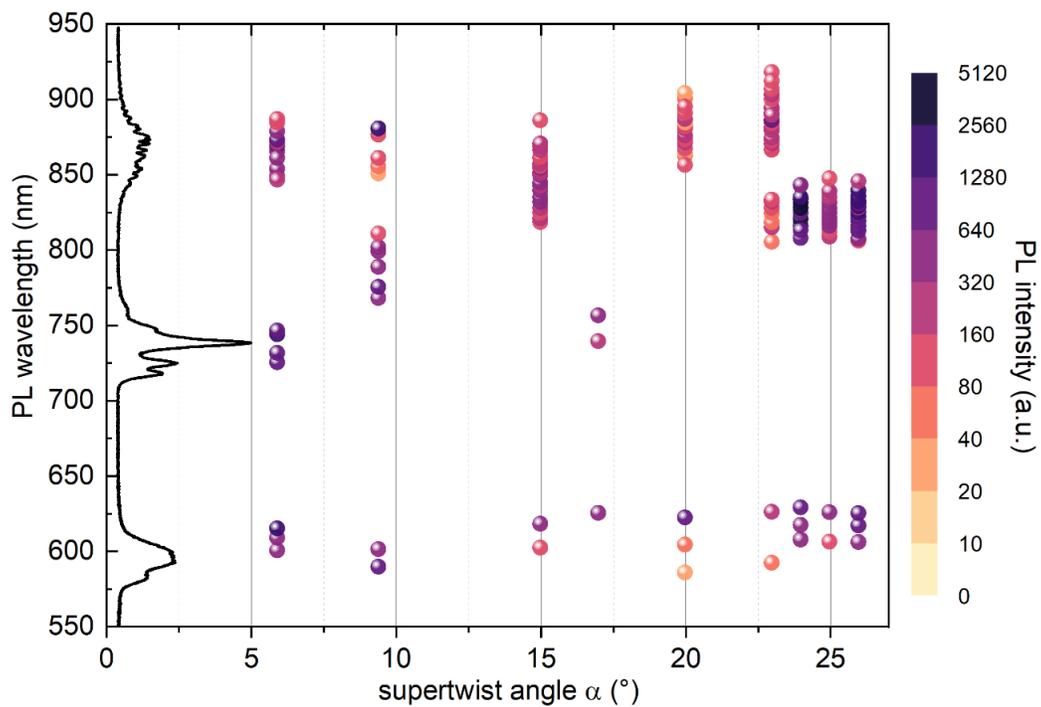

**Fig. S4.** Supertwist angle dependent photoluminescence taken at T=10K. The vertical curve on the left is the PL spectrum for $\alpha = 6°$. The y-coordinates of the data points show the fitted peak positions; the color of the data points represents their intensity. Quantitative prediction of peak position is beyond the scope of this work since the band structure and LDOS calculation methods are not precise in capturing the exact band gap sizes. For large supertwist angles, only two groups of excitons exist indicating that the 3D bandgap might be indirect for these angles.

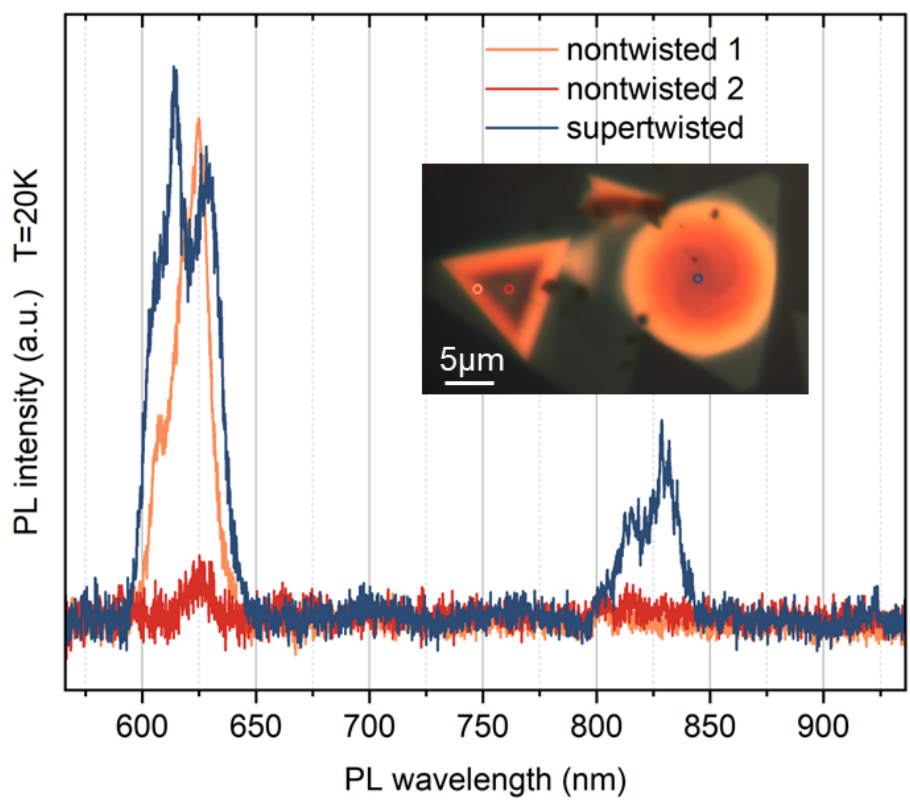

**Fig. S5.** Control group to compare two nearby non-twisted and supertwisted samples The photoluminescence spectra are taken at two supertwisted and nontwisted ceWS2 samples close to each other to show that the emerging exciton peaks are inherited from supertwist structure instead of defects or precursor residues on the substrate. Inset: optical image of the two samples with the position where spectra were taken marked by colored circles with the same color coding as the PL spectra. The red and blue spectra were taken at the centers of nontwisted and supertwisted samples, respectively. The orange curve is taken at the edge of nontwisted sample where the 2D peak has similar brightness with the supertwisted sample, but still shows no 3D peak.

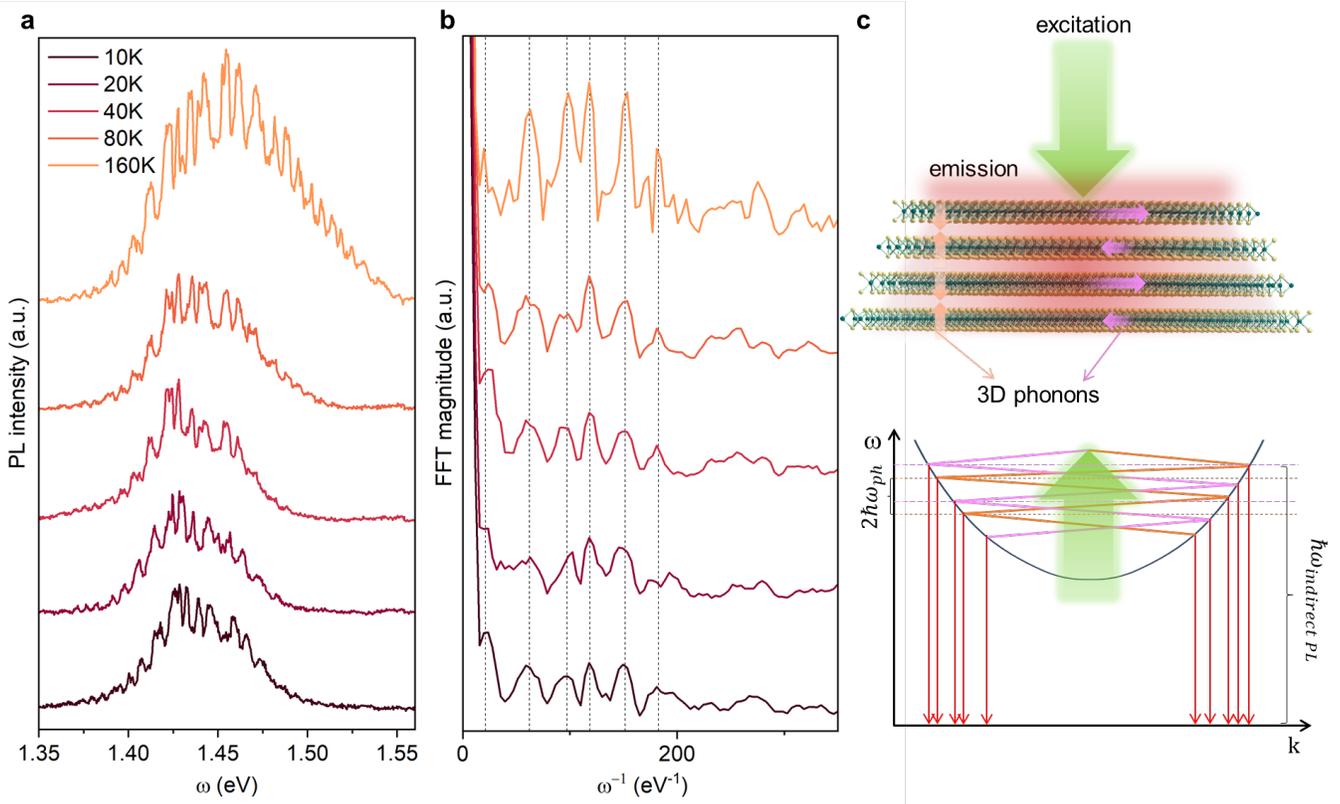

**Fig. S6.** Fourier transform analysis for phonon cascade effect in PL peaks. a. Zoomed indirect 3D excitons from Fig. 2b. The unit of x-axis is changed from wavelength to energy to show the energy spacings between the fine splitting. With increased temperature, these indirect transitions are stronger and broader in energy range. The positions are shifted but the spacing between small peaks are kept as revealed by the fast Fourier transform (FFT) in b. b. FFT of the temperature dependent PL of indirect excitons. The robustness of the FFT peak positions reveals that the peaks in a share the same origin for the spectral splittings. These features can be attributed to 3D breathing and shearing phonon modes in multilayer twisted $WS_2$, and also shift with supertwist angles. c. Schematic of the mechanism for the low-energy phonons to induce fine splitting structure in indirect transitions. Top panel is a real-space structure schematic where orange and pink arrows represent the breathing and shear modes of 3D phonons, and the large green arrows indicate the continuous excitation light of 532 nm. Bottom panel shows a simplified energy band dispersion with phonon cascade after excitation. The low energy phonon modes with energy $\hbar\omega_{ph}$ can be emitted or absorbed before the indirect excitons recombines for emission a series of energies $\hbar\omega_{indirectPL}$.

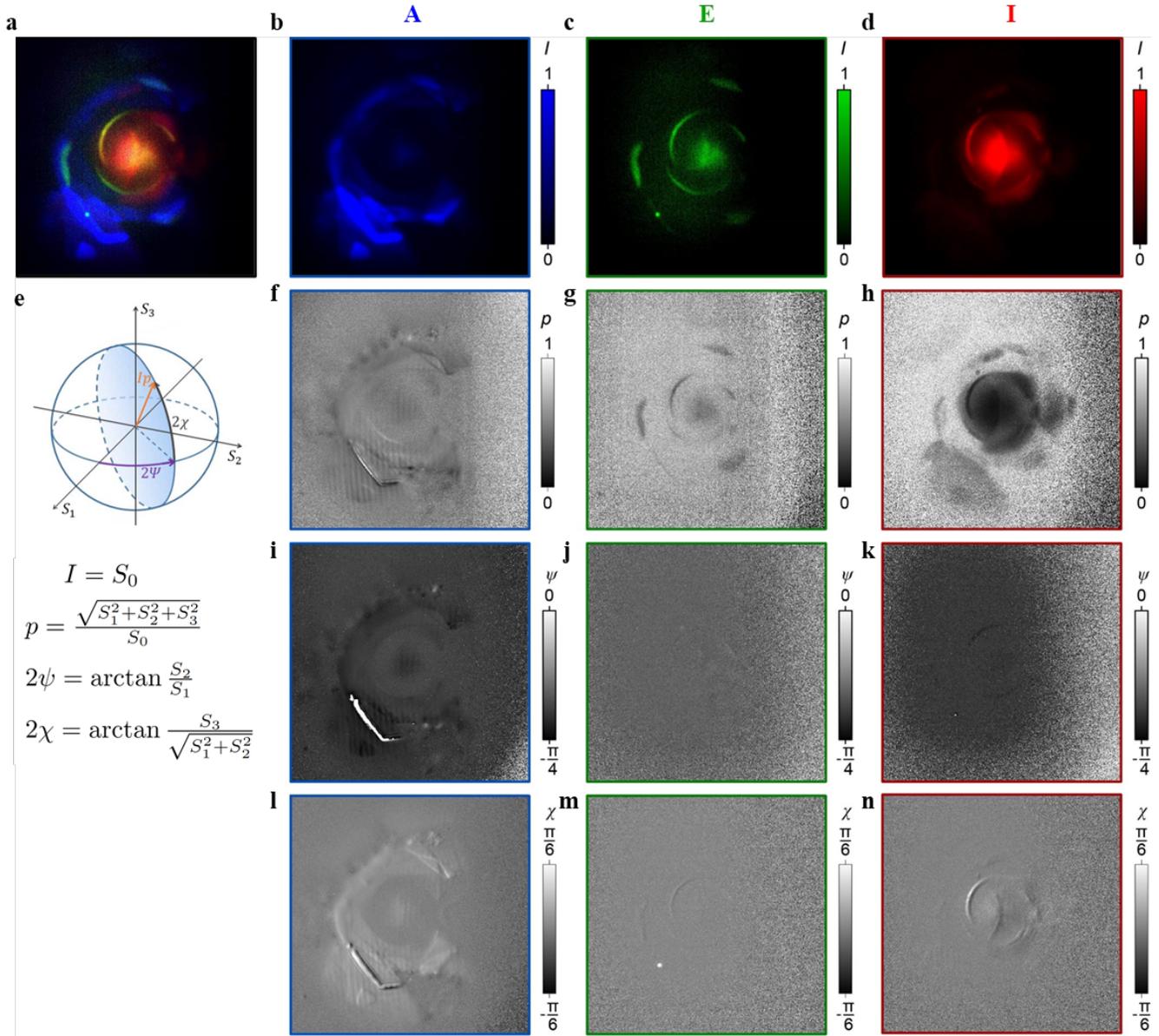

**Fig. S7.** Stokes vector elements (full version of Fig. 3). a. False color intensity map by adding up b-d. b-d, Intensity element maps (I or $S_0$) for A, E, and I excitons, respectively. e. Poincaré sphere illustrating the Stokes representation of the light polarization state. The corresponding relations between Cartesian and spherical coordinates are shown below the sphere. f.-h. p maps for A, E, and I excitons, respectively. i.-k. $\psi$ maps for A, E, and I excitons, respectively. l.-n. $\chi$ maps for A, E, and I excitons, respectively.

**Movies S1 to S6.**

S1-3, Wide-field imaging data of A, E, and I excitons, respectively, as a function of quater-waveplate angles.

S4-6, Local density of states maps of supertwisted WS$_2$ with $\alpha$ =6°, 12°, 24°, respectively, as a function of energy.